# Large Capacity Data Hiding in Binary Image black and white mixed regions

Yuanlin YANG
City University of Hong Kong, China
yuanlyang4-c@my.cityu.edu.hk

*Abstract:* Information hiding technology utilizes the insensitivity of human sensory organs to redundant data, hiding confidential information in the redundant data of these public digital media, and then transmitting it. The carrier media after hiding secret information only displays its own characteristics, which can ensure the transmission of confidential information without being detected, thereby greatly improving the security of the information. In theory, any digital media including image, video, audio, and text can serve as a host carrier. Among them, hiding information in binary images poses great challenges. As we know, any information hiding method involves modifying the data of the host carrier. The more information hidden, the more data of the host carrier are modified. In this paper, we propose information hiding in the black-and-white mixed region of binary images, which can greatly reduce visual distortion. In addition, we propose an efficient encoding to achieve high-capacity information hiding while ensuring image semantics. By selecting binary images of different themes, we conduct experiments. The experimental results prove the feasibility of our technique and verify the expected performance. Since the candidate units for information hiding are selected from equally sized blocks that the image is divided into, and the hiding and extraction of information are based on a shared encoding table, the computational cost is very low, making it suitable for real-time information hiding applications.

*Keywords: Information hiding, Encoding, Decoding, Image processing.*

## I. Introduction

Information hiding, also known as data hiding, is an important field of technology that combines multidisciplinary theory and technique [1, 2]. It mainly refers to embedding specific information in the digital host digital media, such as text, digitized sound, image, video signal, etc. [3-7]. The purpose of information hiding is not to restrict normal information access and access, but to ensure that the hidden information does not attract the attention of a third party or the monitor, to reduce the possibility of attack and detection. In contrast, cryptography is used to strengthen the security of information by disordering the data of information itself. It should be noted that cryptography and information hiding technologies are not contradictory and competing, but complementary. The difference between them is that the application occasions and requirements for the algorithm are different, but they may cooperate in practical applications [8, 9]. According to occasions, information hiding is also named steganography and digital watermarking Steganography involves embedding secret information into seemingly ordinary messages to prevent third parties from detecting the secret information [10, 11]. Digital watermarking is to have a specific identifiable meaning of the mark (watermark) permanently embedded in the host media, mainly copyright-protected digital works, and will not affect the quality of the host data. Digital watermarking technology is mainly used for copyright protection, copy control, and operation tracking [12]. In practice, all digital media, such as images, audio, video, and text, can be used as the carrier of information hiding.

The essence of information hiding is to make use of the data redundancy of host media, and the insensitivity of human eyes and ears to certain modifications of host media redundant data. Based on self-defined modification rules, information is hidden in the host media, and people cannot perceive the existence of hidden information in the media. In order words, it uses people's WYSIWYG (What You See Is What You Get) perception to confuse attackers. Therefore, it is important to understand the characteristics of common media. In short, the redundant data of the host media acts as the carrier of the information to be hidden and should be modified, if applicable, to meet the hiding rules together with the information to be hidden.

The difficulty of information hiding lies in how to inject a large amount of information into a given host media, while the data of the host media has only changed slightly to make it difficult for human eyes to detect the abnormality of the host media. As an example, a color image is rich in color range, and its depth of gray is also in a long range, so it is easier to hide information in a color image [13]. In contrast, it is very difficult to hide information in a binary image, because any change in the pixel value of a binary image corresponds to a significantly different brightness [14]. Therefore, data hiding in binary images is a challenging task. In this paper, we choose binary images as host media and propose a novel technique to achieve high-capacity data hiding in them.

## II. Analysis of binary images

In a binary image shown in Fig. 1, a pixel value is either '1' or '0' indicating white or black respectively. If a pixel value needs to be changed, or flipped, due to data hiding, it will cause a white dot ('1") to become a black one or a black dot to become a white dot. Therefore, data hidden in binary images typically causes greater visual distortion than in grayscale or color images.

To avoid this, we propose rules for slightly modifying the edge regions of binary images, as shown in Fig. 2. In an edge region with fixed size, if the number of black dots is adjusted from 1 or 3 to 2, the distortion in that area will not be too severe. Similarly, if the number of black dots is adjusted from 2 to 1 or 3, the distortion in that area will not be too severe.

Generally, at the edge of a closed black area, a small increase or decrease in black spots is not easily detected by the human eye. Similarly, at the edge of a closed white area, a small increase or decrease in white spots is imperceptible to the human eyes. We set up rules for slightly modifying the edge regions of binary images to implement data hiding. For example, for an edge block of size 2x2, to make it recognizable on the receiving end, we change the number of block pixels from odd (1 or 3) to even (2), or from even (2) to odd (1 or 3), shown in Fig. 2. Using this method will effectively reduce visual distortion caused by data hiding. The detailed rules will be introduced in Section 3.

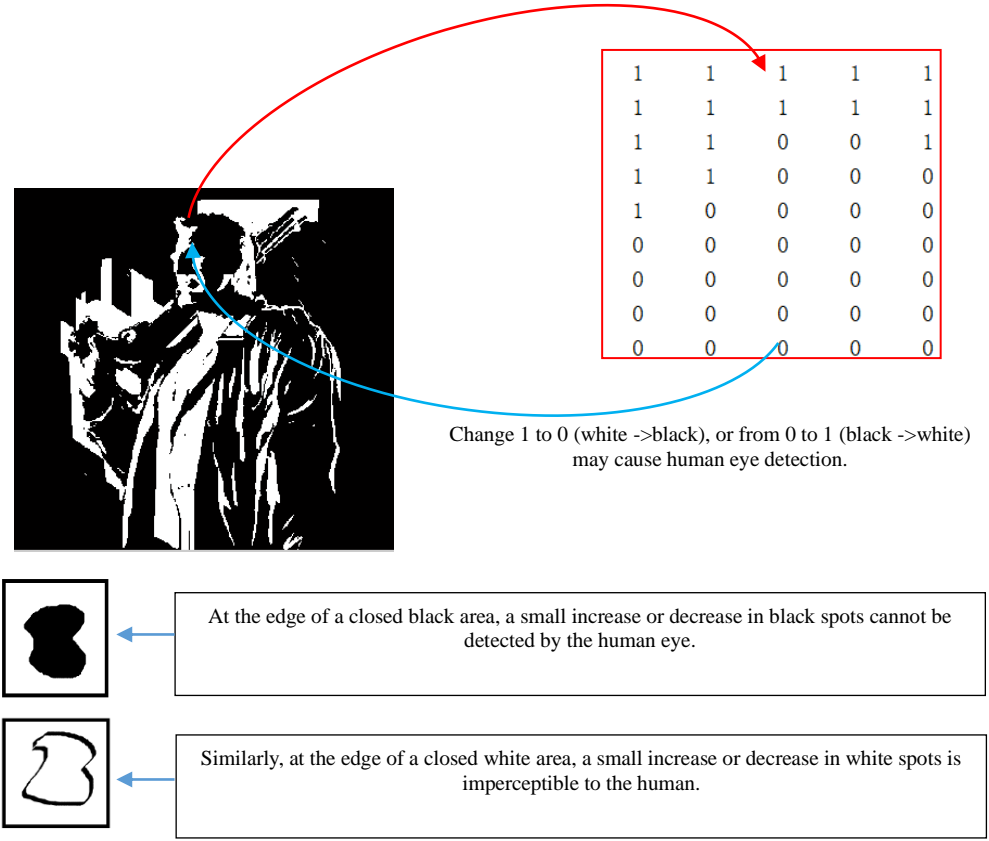

Fig. 1 Analysis of binary image

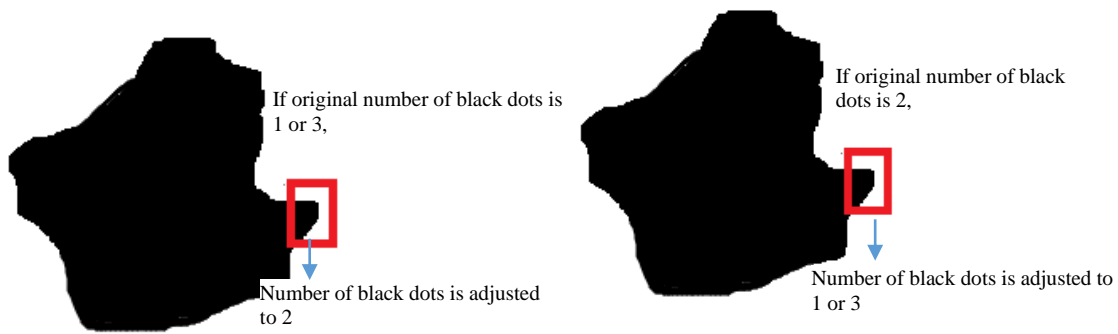

Fig. 2 Rules for slightly modifying the edge regions of binary images

### III. Proposed data hiding technique in binary images

Data hiding work consists of two parts, namely, data hiding and data extraction which is the reverse of data hiding. In our proposed technique, to reduce the distortion caused by hiding information in it, we hide the information in suitable image blocks. In other words, the basic unit carrier of data hiding is a suitable image block.

#### A. Image blocking

We divide binary images into equal-sized non-overlapping blocks as shown in Fig. 3. Assume the size of the image is

$MXN$, and the size of a block is $k$x$k$. Then the number of blocks we can get is as follows:

$$m=\lfloor M/k \rfloor, n=\lfloor N/k \rfloor \quad (1)$$

where $\lfloor . \rfloor$ is a rounding-down operation. In (1), $m$ is the number of blocks in the row direction, and $n$ is the number of blocks in the column direction. The total number of blocks is $m*n$.

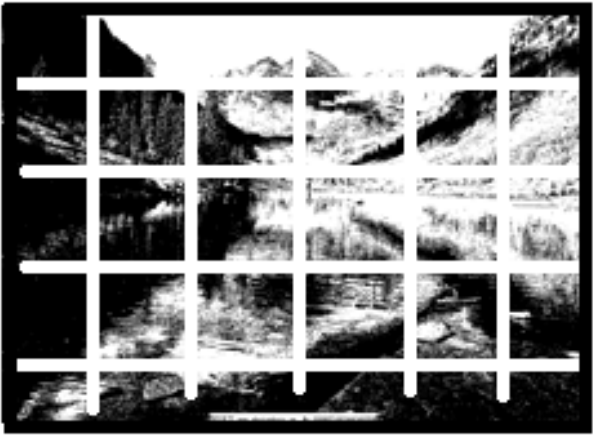

Fig. 3 Example of image blocking

For a given image, the larger the value of $k$, the less the total number of blocks we can get; the smaller the value of $k$, the more the total number of blocks we can get. If the block is too large, the data hiding capacity is small. If the block is too small, it is difficult to hide the data within it without causing significant semantic distortion. In this paper, we choose $k=2$, to make a tradeoff between capacity and semantic distortion.

*B. Block selection and data hiding rule*

For a block, there are $2^4$ different patterns, i.e., black-white dot combinations, in total. Pure white or pure black blocks are not suitable for hiding data, because any flipping of pixel values (0->1 or 1->0) causes too obvious visual distortion. The remaining 14 (or, $2^4 -2$) types of blocks are classified into two categories:

A: There are 2 black pixels in the block;

B: There are 1 or 3 black pixels in the block.

Each block belonging to A or B is used to hide information, but the methods are slightly different.

**For blocks belonging to A**

For a block belonging to A, we hide 3 bits of information by replacing 4-pixel values of the block with 3-bit information, complying with Table 1. After data hiding, the block contains either 1 or 3 black dots. For example, if the 3 bits to be hidden is '011', the host block belonging to A is forcibly changed to '0111'. ( '0' refers to a black pixel value, '1' refers to a white pixel value. '0111' means the 1st row of the block is '01', and the 2nd row is '11').

Table 1 Data hiding code table I

| 3 bits of information to be hidden | The pattern of the host block changed due to data hiding |
|---|---|
| 000 | 0001 |
| 001 | 0010 |
| 010 | 0100 |
| 011 | 0111 |
| 100 | 1000 |
| 101 | 1001 |
| 110 | 1101 |
| 111 | 1110 |

To extract information from the block, we just check whether it contains 1 or 3 black dots. If yes, 3 bits of information can be extracted by reversely looking up Table 1.

**For blocks belonging to B**

For a block belonging to B, we hide 2 bits of information by replacing 4-pixel values of the block with 2-bit information, complying with Table 2. After data hiding, the block contains 2 black dots.

Table 2 Data hiding code table II

| 2 bits of information to be hidden | The pattern of the host block changed due to data hiding |
|---|---|
| 00 | 0011 |
| 01 | 0101 |
| 10 | 0110 |
| 11 | 1001 |

To extract information from the block, we just check whether it contains 2 black dots. If yes, 2 bits of information can be extracted by reversely looking up Table 2.

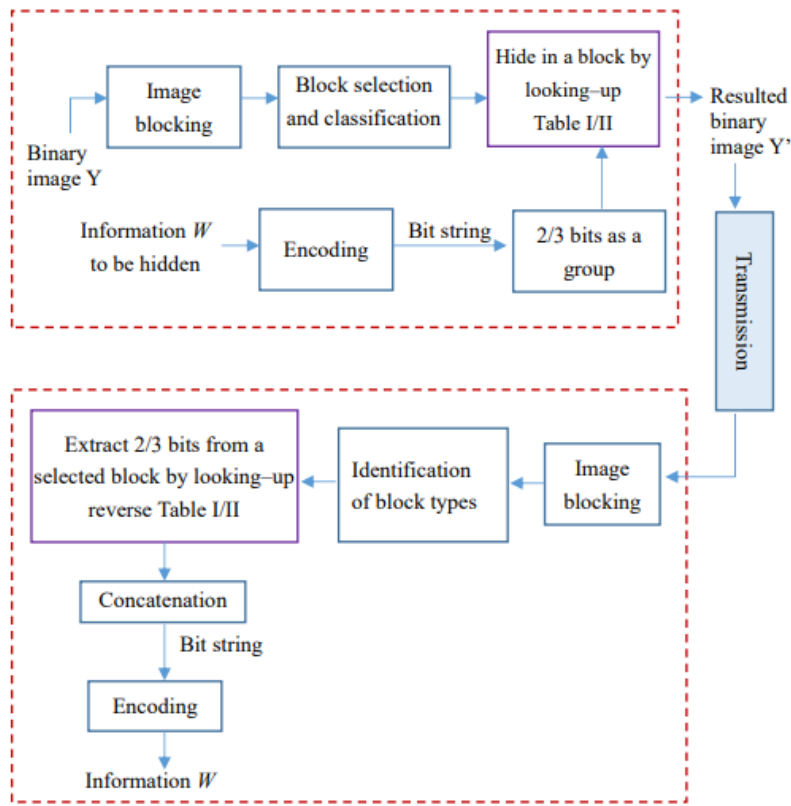

Fig. 4 Overall scheme of the proposed technique

The overall scheme of the proposed technique is shown in Fig. 4. First of all, take a binary image Y1 as host media. Divide it into 2*2 blocks, then screen all blocks one by one, skipping pure white or pure black blocks. The remained blocks are used for data hiding. Based on the category to which a block belongs, we hide 3 bits or 2 bits in a block. After all bits are hidden, we get the resulting image Y1'. This completes the data hiding process, as shown in the dashed box at the top in Fig. 4. (For simplicity, we assume there are enough blocks available for data hiding).

When Y1' arrives at the destination, the receiver will extract information from it. The data extraction process is shown in the dashed box at the bottom of Fig. 4.

IV. Experimental results and analysis

To reflect the authenticity of the experiment, we randomly selected 6 binary images with different themes. To facilitate comparison, we deliberately adjust them to the same resolution (1024*768), as shown in Fig. 5. We use MATLAB R2021 to implement our algorithms, and execute the program on PC with CPU 7950X3D, RAM 32G.

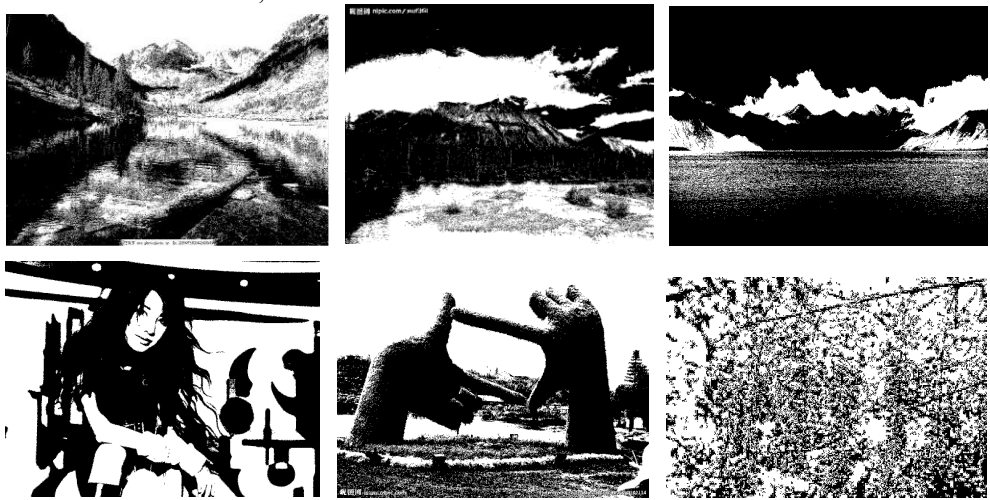

Fig. 5 Original host images used (first row: 1-3, second row: 4-6)

## A. Capacity

The number of bits that we can completely hide in each host image is recorded in Table 3.

Table 3 Data hiding capacity

| Images | 1# | 2# | 3# | 4# | 5# | 6# |
|---|---|---|---|---|---|---|
| Capacity (bits) | 105085 | 44664 | 52482 | 14132 | 57667 | 64069 |

From Table 3, we can see that the capacity is different, though the size of the images is the same. The capacity of Image No. 1 is the highest, and that of Image No. 4 is the lowest. This is because there are few areas (blocks) of pure white or pure black blocks in Image No. 1, while there are too many in Image No. 4. Pure white or pure black blocks are not used to hide data. The larger the size of an image of the same theme, the more information can be embedded. The relevant experimental data is not included in this table. Figure 6 shows the resulting images with data hiding capacity of Table 3. In Fig. 6, no obvious distortion in the theme and semantics of the host images can be seen.

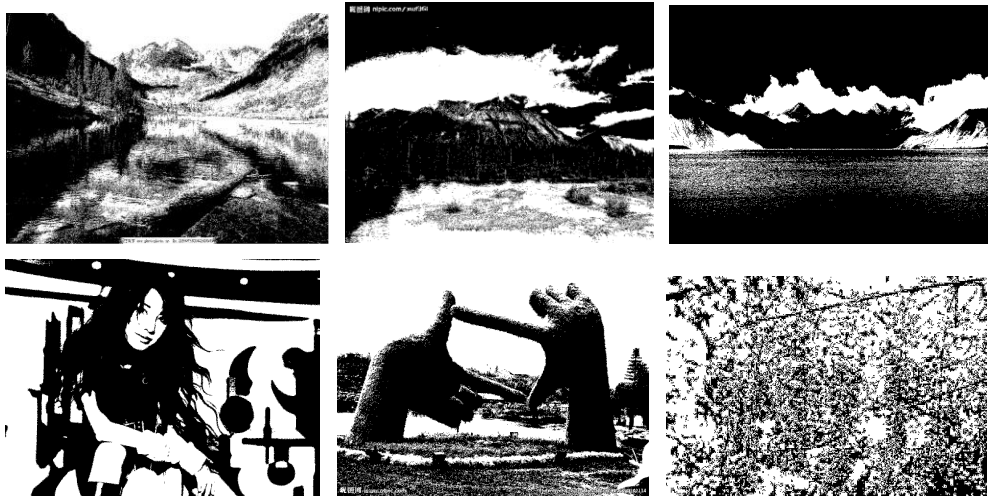

Fig. 6 Resulted images after data hiding (first row: 1-3, second row: 4-6)

## B. PSNR

To objectively evaluate the overall degree of changes in image data caused by information hiding, we adopt the PSNR formula. In the communication system, PSNR is an engineering term that represents the ratio of the maximum possible power of a signal to the destructive noise power that affects its representation accuracy. When applied to images, PSNR is one of the indicators used to measure image quality. It is based on the definition of MSE (Mean Square Error), which can be defined as:

$$MSE = \frac{1}{mn} \sum_{i=0}^{m-1} \sum_{j=0}^{n-1} [I(i,j) - K(i,j)]^2$$

$$PSNR = 10 \cdot \log_{10} \left( \frac{MAX_I^2}{MSE} \right) = 20 \cdot \log_{10} \left( \frac{MAX_I}{\sqrt{MSE}} \right)$$

We use PSNR to evaluate the host image's distortion due to data hiding. Therefore, the I(i, j) and K(i, j) in the formula are the original host image and the resulting image after data hiding, respectively. We conducted the experiments and recorded the results in Table 4.

Table 4 Data hiding PSNR

| Images | 1# | 2# | 3# | 4# | 5# | 6# |
|---|---|---|---|---|---|---|
| PSNR (dB) | 37.6340 | 41.3097 | 40.6804 | 46.5062 | 40.1598 | 39.9050 |

From Table 4, we find the PSNR values are different, though the size of the images is the same. The PSNR of Image No. 1 is the lowest, and that of Image No. 4 is the highest. This is the number of bits we hide in Image No. 1 is the least, while the most in Image No. 4. For images with the same size, the more information bits embedded, the lower the fidelity of the image and the smaller the PSNR. This further confirms that the capacity of information hiding conflicts with the fidelity of the carrier.

## C. Running time

Due to the different running states of the computer in different periods, there will be some deviation in the recorded running time. To minimize the occurrence of such deviation, we averaged multiple runs and listed the results in Table 5.

Table 5 Data hiding execution time

| Host images | 1# | 2# | 3# | 4# | 5# | 6# |
|---|---|---|---|---|---|---|
| Hiding time (ms) | 400 | 87 | 156 | 0.7 | 158 | 189 |
| Extraction time(ms) | 699 | 294 | 400 | 186 | 364 | 413 |

From Table 5, it can be seen that our data retention time and data transmission time are both very short. This is because our algorithms do not involve complex mathematical operations, just basically looking up the linear table of small size. In the case of the same computer performance, the larger the host image, the more blocks divided, then the longer the time it takes to run. The length of time for hiding and extracting is also related to the number of valid blocks.

## V. Conclusions

Black and white binary image has the advantages of a small amount of data and high transmission efficiency. However, since black and white binary image is only composed of two types of pixels, either white or black. Compared with data hiding in color images or gray images, data hiding in binary images with a large amount of information is very challenging. In this paper, we propose an adaptive binary image data hiding technique. A binary image is pre-divided into equal-sized blocks, and only suitable blocks are selected as basic carriers of data hiding to reduce visual distortion due to data hiding. The proposed technique can achieve the high capacity of data hiding and low semantic distortion of the resulting image. Since the proposed algorithms are simple, the running times of both data hiding and data extraction in an ordinary PC are very short. Thus, the technique can be used in real-time data hiding applications. Numerous experimental results have confirmed the effectiveness of our method. In the future, we will consider hiding data in blocks of varying sizes or even irregular shapes.